\documentclass[12pt]{article}
\usepackage[english]{babel}
\usepackage{amsmath}
\usepackage{amssymb}
\usepackage{graphicx}
\usepackage[mathscr]{eucal}%
\usepackage{cite}
\usepackage{amsfonts}
\begin{document}

\title{
\bf{}On Lagrangian Formulation for Half-integer  HS Fields within
Hamiltonian BRST Approach }
\author{A.A.
Reshetnyak\thanks{reshet@tspu.edu.ru} \\[0.5cm]
\it Laboratory of Nonlinear Media Physics, \it   Institute of
Strength Physics\\ and Materials Science, \it Tomsk 634021,
Russia}
\date{}

\maketitle

\begin{abstract}
A recent progress in a gauge-invariant Lagrangian description
of massive and massless half-integer higher-spin fields in AdS
and Minkowski spaces is presented. The procedure is based on
a BFV-BRST operator, encoding the initial conditions realized
by constraints in a Fock space and extracting the higher-spin
fields from unitary irreducible representations
of the AdS (Poincare) group. The construction is applicable to
higher-spin tensor fields with a multi-row Young tableaux.
\end{abstract}


\paragraph{1. Introduction}Problems of higher-spin (HS) field
theory remain an important area of scientific research in view of
their close relation to superstring theory, which operates with
an infinite set of bosonic and fermionic HS fields, including
massless and massive fields (for a review, see \cite{reviews}).
This article takes a snapshot of constructing a Lagrangian
formulation (LF) for free half-integer HS fields as a starting
point for an interacting HS field theory in the framework of
conventional Quantum Field Theory, and is based on the results
presented in \cite{adsfermBKR,flatfermmix}.

The methods of constructing an LF for HS fields are based on the
BFV--BRST approach \cite{BFV}, developed in a way that applies to
Hamiltonian quantization of gauge theories with a given LF, and
consists in a solution of the problem inverse to that of the
method \cite{BFV} (as in the case of string field theory
\cite{SFT} and in the early papers on HS fields \cite{Ouvry}) in
the sense of constructing a gauge LF with respect to a nilpotent
BFV--BRST operator $Q$. This operator, in its turn, is constructed
from a system $O_\alpha$ of first-class constraints defined in an
auxiliary Fock space and encoding the relations that extract the
fields with a definite mass and spin from the representation
spaces of the AdS or Poincare group.

\vspace{-1ex} \noindent
\paragraph{2. Fermionic fields in AdS spaces} It is well-known that massive half-integer spin
$s=n+\frac{1}{2}$
representations of
the AdS group are realized in the space of totally symmetric
tensor-spinor fields $\Phi_{\mu_1\ldots\mu_n}(x)$, with suppressed
 Dirac index, satisfying the equations
\begin{eqnarray}
\label{Eq-00} && \bigl[i\gamma^{\mu}\nabla_{\mu}
    -r^\frac{1}{2}(n+\textstyle\frac{d}{2}-2)-m
\bigr] \Phi_{\mu_1\mu_2\ldots\mu_n}(x) =0,\qquad  \gamma^{\mu}
\Phi_{\mu\mu_2\ldots\mu_n}(x) =0.\footnotemark
\end{eqnarray}\footnotetext{Where $r$ is the radius of a d-dimensional AdS space $r=\frac{R}{d(d-1)}$,
 with $R$
being the scalar curvature, $g_{\mu\nu}$ having the mostly minus
signature, and Dirac's matrices satisfying the relation
$\bigl\{\gamma_\mu,\gamma_\nu\bigr\}=2g_{\mu\nu}$.}For a simultaneous
description of all half-integer HS fields,  one introduces a Fock
space $\mathcal{H}$ generated by creation $a^+_\mu(x)$ and
annihilation $a_{\mu}(x)$ operators, $[a_\mu, a_\nu^+] =
-g_{\mu\nu}$, and a set of  constraints for an arbitrary vector
$|\Phi\rangle \in \mathcal{H}$,
\begin{eqnarray}
\label{t't0}  {\tilde{t}}'_0|\Phi\rangle & = &
\Bigl(-i\tilde{\gamma}^\mu D_\mu + \tilde{\gamma}\left( m +
r^{\frac{1}{2}} (g_0 - 2)\right)\Bigr)|\Phi\rangle=0 ,\qquad
{t}_1|\Phi\rangle = \tilde{\gamma}^\mu a_\mu |\Phi\rangle=0,\\
\label{PhysState}  |\Phi\rangle & = &
\sum_{n=0}^{\infty}\Phi_{\mu_1\ldots\,\mu_n}(x)\,
a^{+\mu_1}\ldots\,a^{+\mu_n}|0\rangle,
\end{eqnarray}
given in terms of an operator $D_\mu$ equivalent to $\nabla_{\mu}$
in its action in $\mathcal{H}$, as well as in terms of fermionic operators
${\tilde{t}}'_0, {t}_1$ constructed from an enlarged
set of Grassmann-odd gamma-matrix-like objects
$\tilde{\gamma}^\mu, \tilde{\gamma}$
($\{\tilde{\gamma}^\mu,\tilde{\gamma}^\nu\} = 2\eta^{\mu\nu}$,
$\{\tilde{\gamma}^\mu,\tilde{\gamma}\}=0$, $\tilde{\gamma}^2=-1$
\cite{adsfermBKR, flatferm}), related to conventional gamma-matrices
as $\gamma^{\mu} = \tilde{\gamma}^{\mu} \tilde{\gamma}$.
The validity of relations (\ref{t't0}) is equivalent to the
simultaneous fulfilment of Eqs. (\ref{Eq-00}) for all fermionic
HS fields $\Phi_{\mu_1\ldots\,\mu_n}(x)$.

To obtain a Hermitian BFV--BRST charge, it is necessary to deduce
a set of first-class quantities $O_I: \{O_\alpha\} \subset
\{O_I\}$ closed under the operations of a) Hermitian conjugation
with respect to an odd scalar product
$\langle\tilde{\Psi}|\Phi\rangle$ \cite{adsfermBKR} and b)
supercommutator multiplication $[\ ,\ \}$. As a result of the
first step in obtaining the set of $\{O_\alpha\}$, the final
original massive half-integer HS symmetry superalgebra, $\{o_I\}$
= $\{{\tilde{t}}'_0, {t}_1, {t}_1^+, {l}_i, {l}_i^+, g_0,
\tilde{l}_0'\}$, $i=1,2$,
\begin{align}
& (t_1^+,g_0) = (\tilde{\gamma}^\mu a_\mu^+, -a^+_\mu
a^\mu+{\textstyle\frac{D}{2}}) ,\  (l_1, l_1^+) =-i(a^\mu,
a^{+\mu})D_\mu ,\
(l_2,l_2^+)={\textstyle\frac{1}{2}}(a^\mu a_\mu, a^{+\mu}a^+_\mu),
\end{align}
\vspace{-3em}
\begin{eqnarray}
{\tilde{l}}'_0=g^{\mu\nu}(D_\nu
D_\mu-\Gamma^\sigma_{\mu\nu}D_\sigma)
-r\left(g_0+t_1^+t_1+{\textstyle\frac{d(d-3)}{4}}\right) + \left(m
+ r^{\frac{1}{2}} (g_0 - 2)\right)^2, \label{l'l0}
\end{eqnarray}
contains a central charge $\tilde{m}=(m-2\sqrt{r})$, a subset of
6 second-class constraints $\{o_a\} = \{t_1,t^+_1, l_i, l_i^+\}$,
a quantity $g_0$, composing, together with $\tilde{m}$, an invertible
supermatrix $\|[o_a, o_b\}\|$, and satisfies some quadratic
algebraic relations.

To convert the subsystem $o_a$ into the first-class system $O_a$,
we apply an additive\nolinebreak{\;}con\-ver\-sion procedure for
nonlinear superalgebras developed in \cite{myreport}, which
consists, first of all, in constructing additional (with respect
to $o_I$) parts $o'_I$ acting in a new Fock
space\nolinebreak{\,}$\mathcal{H}'$ generated by fermionic $f,f^+$
and bosonic $b_i, b_i^+$, $i=1,2$, creation and annihilation
operators, so that the converted constraints $O_I = o_I + o'_I$
satisfy\nolinebreak{\,}a\nolinebreak{\,}new\nolinebreak{\,}algebra:\nolinebreak{\,}$[O_I,
O_J\} \sim O_K$.

The condition of additivity,  $[o_I,o'_I\}=0$, cannot be fulfilled
for $o_I$ due to the presence of the $\tilde{\gamma}$-matrix in its
definition. Therefore, we pass to another basis of constraints, $o_I
\to \tilde{o}_I = u^J_Io_J$, by means of a nondegenerate
transformation, such that only ${\tilde{t}}'_0, {\tilde{l}}'_0$ are
changed, ${{t}}_0 = -i\tilde{\gamma}^\mu D_\mu, l_0 = -t_0^2$, and
such that after the additive conversion $\tilde{O}_I = \tilde{o}_I + o'_I$
we can make an inverse transformation ($u^{-1}$ being enlarged in
$\mathcal{H}\bigotimes \mathcal{H}'$ to $U^{-1}$) of the converted
constraints $\tilde{O}_I$ to ${O}_I = (U^{-1})^J_I\tilde{O}_I $.
Next, the construction of the additional parts $o'_I$ is based on
the condition $[\tilde{O}_I, \tilde{O}_J\} \sim \tilde{O}_K$
that implies an unambiguous form of the superalgebra of $\{o'_I\}$:
\begin{eqnarray}
[\,o_i',o_j'\}_s &=&
f_{ij}^ko_k'-(-1)^{\varepsilon(o_k)\varepsilon(o_m)}f_{ij}^{km}o_m'o_k'
, \label{addal}
\end{eqnarray}
(with the Grassmann parity $\varepsilon(o_m) = 0,1$ respectively
for bosonic and fermionic $o_m$), provided that the form of one of
$\{o'_I\}$ is given by $[\tilde{o}_i,\tilde{o}_j\}_s =
f_{ij}^k\tilde{o}_k+f_{ij}^{km}\tilde{o}_k\tilde{o}_m$
\cite{adsfermBKR}. Then, the enlarged central charge $\tilde{M} =
\tilde{m} + \tilde{m}'$ vanishes, whereas the
 explicit expressions for $o'_I$ in terms of the operators $f,f^+, b_i, b_i^+$ and
  new constants $m_0, h$ [they are to be determined later from
the  condition  of reproducing the correct form of
Eqs.(\ref{t't0})] are presented in \cite{adsfermBKR} and  can be
found following the method described in \cite{flatferm} for
totally-symmetric half-integer HS fields in flat spaces, as well
as in \cite{0206027} for massive integer HS fields in AdS spaces
extended to the case of the Verma module construction for
nonlinear superalgebras.

We then proceed to construct a BFV-BRST operator $\tilde{Q}'$ for
the system of operators $\tilde{O}_I$ in the case of the Weyl ordering
for quadratic combinations of $\tilde{O}_I$ in the right-hand sides of
$[\tilde{O}_I, \tilde{O}_J\}$ and for the $(\mathcal{C}\mathcal{P})$-ordering
for the ghost coordinates $\mathcal{C}^I$: bosonic $q_0,
q_1,q_1^+$ and fermionic $\eta_0$, $\eta_1^+$, $\eta_1$,
$\eta_2^+$, $\eta_2$, $\eta_G$, and their conjugated momenta
$\mathcal{P}_I$: $p_0$, $p_1^+$, $p_1$, ${\cal{}P}_0$,
${\cal{}P}_1$, ${\cal{}P}_1^+$, ${\cal{}P}_2$, ${\cal{}P}_2^+$,
${\cal{}P}_G$, with the standard ghost number distribution
$gh(\mathcal{C}^I)$ = $ - gh(\mathcal{P}_I)$ = $1$, providing
$gh(\tilde{Q}')$ = $1$. In contrast to bosonic HS fields,  in
the given basis the operators $\{\tilde{O}_I\}$ form an open algebra
with respect to $[\ ,\ \}$ ($[\tilde{O}_I , \tilde{O}_J\}$ =
$F_{IJ}^K(\tilde{O}, o')\tilde{O}_K$), so that the nilpotent operator
$\tilde{Q}'$ corresponds to a formal second-rank topological gauge theory,
\begin{equation}\label{generalQ'}
    \tilde{Q}'  = {O}_I\mathcal{C}^I + \frac{1}{2}
    \mathcal{C}^{I_1}\mathcal{C}^{I_2}F^J_{I_2I_1}\mathcal{P}_J (-1)^{\varepsilon({O}_{I_2} +
    \varepsilon({O}_J)}+\frac{1}{6}\mathcal{C}^{I_1}\mathcal{C}^{I_2}\mathcal{C}^{I_3}
    F^{J_2J_1}_{I_3I_2I_1}\mathcal{P}_{J_2}\mathcal{P}_{J_1}
\end{equation}
with completely definite functions
$F^{J_2J_1}_{I_3I_2I_1}(\tilde{O}, o')$ resolving the Jacobi
identity for $\tilde{O}_I$ \cite{adsfermBKR, myreport}.

A covariant extraction of the operator $G_0$ from the system
$\{\tilde{O}_I\}$, in order to pass to the converted first-class
constraints $\{\tilde{O}_\alpha\}$ only, is based on the condition of
independence of  $\mathcal{H}_{tot} \equiv \mathcal{H}\bigotimes
\mathcal{H}'\bigotimes \mathcal{H}_{gh}$ of
$\eta_G$\nolinebreak{\,}and\nolinebreak{\,}on the elimination\nolinebreak{\,}
from\nolinebreak{\,}$\tilde{Q}'$ of the terms proportional to
$\eta_G, \mathcal{P}_G$\nolinebreak{\,}\cite{adsfermBKR}:
\begin{eqnarray}
\tilde{Q}' = \tilde{Q} + \eta_G(\sigma+h)+\mathcal{B}
\mathcal{P}_G,\quad \sigma+h = g_0+g'_0 + \left(iq_1^+p_1 +
\textstyle\sum_{k=1}^2k\eta^{k+}\mathcal{P}_k + h.c.\right);
\label{decomposQ'}
\end{eqnarray}
the same applies to the physical vector $|\chi\rangle \in
\mathcal{H}_{tot}$, $|\chi\rangle = |\Phi\rangle +
|\Phi_A\rangle$, $|\Phi_A\rangle_{\{b_i = b_i^+ = f = f^+ =
\mathcal{C} = \mathcal{P} = 0\}}$ = $0$, with the use of the
BFV-BRST equation $\tilde{Q}'|\chi\rangle = 0$ determining the
physical states:
\begin{align}
\label{Qchi} & \tilde{Q}|\chi\rangle=0\footnotemark, &&
(\sigma+h)|\chi\rangle=0, && \left(\varepsilon,
{gh}\right)(|\chi\rangle)=(1,0).
\end{align}\footnotetext{In \cite{adsfermBKR, myreport} it is shown that
the transition $\tilde{O}_I \to O_I$ is realized in $\mathcal{H}_{tot}$ by means
of a unitary transformation constructed with respect to
$\|U^{-1}\|$, so that $\tilde{Q} = Q$ and ${Q}^2|\chi\rangle=0$ if
$(\sigma + h)|\chi\rangle=0$.}Note that the second equation must
take place in the entire $\mathcal{H}_{tot}$, thus determining the spectrum
of spin values for $|\chi\rangle$, whereas the first equation is valid only in
the subspace of $\mathcal{H}_{tot}$ with the zero ghost number.

The presence of an extensive gauge ambiguity in the definition of an LF
permits a covariant separation in $Q$ of all the operators with second-order
derivatives with respect to $x^\mu$, thus expanding $Q$ in
the powers of the zero-mode pairs $q_0, p_0, \eta_0, \mathcal{P}_0$ as
follows \cite{adsfermBKR}:
\begin{eqnarray}
\label{strQ} Q&=&q_0\tilde{T}_0+\eta_0\tilde{L}_0
+i(\eta_1^+q_1-\eta_1q_1^+)p_0 +(q_0^2-\eta_1^+\eta_1){\cal{}P}_0
+\Delta{}Q.
\end{eqnarray}
As a result, due to the representation $|\chi\rangle
 =\sum_{k=0}^{\infty}q_0^k( |\chi_0^k\rangle
+\eta_0|\chi_1^k\rangle)$, the first equation in (\ref{Qchi})
takes the form
\begin{eqnarray}
 \Delta{}Q|\chi^{0}_{0}\rangle
+\frac{1}{2}\bigl\{\tilde{T}_0,\eta_1^+\eta_1\bigr\}
|\chi^{1}_{0}\rangle =0, \qquad \tilde{T}_0|\chi^{0}_{0}\rangle +
\Delta{}Q|\chi^{1}_{0}\rangle =0, \label{EofM2all}
\end{eqnarray}
which can be deduced from the Lagrangian action
\begin{eqnarray}
{\cal{}S}=
\langle\tilde{\chi}^{0}_{0}|K\tilde{T}_0|\chi^{0}_{0}\rangle +
\frac{1}{2}\,\langle\tilde{\chi}^{1}_{0}|K\bigl\{
   \tilde{T}_0,\eta_1^+\eta_1\bigr\}|\chi^{1}_{0}\rangle
+ \langle\tilde{\chi}^{0}_{0}|K\Delta{}Q|\chi^{1}_{0}\rangle +
\langle\tilde{\chi}^{1}_{0}|K\Delta{}Q|\chi^{0}_{0}\rangle.
\label{L1}
\end{eqnarray}
In (\ref{L1}), we have used an odd scalar product in
$\mathcal{H}_{tot}$ and a nondegenerate operator $K =
\hat{1}\bigotimes K'\bigotimes \hat{1}_{gh}$, which provides the
Hermitian character of the operators with respect to $\langle\
|\ \rangle$, as well as the reality of ${\cal{}S}$ (for details,
see \cite{adsfermBKR}). The corresponding LF of a HS field with
a specific value of spin $s=n+\frac{1}{2}$ is a reducible gauge
theory of $L = n-1$-th stage of reducibility.

 \vspace{-1ex} \noindent
\paragraph{3.
Fermionic fields with an arbitrary Young tableaux} Let us
examine the construction of an LF for spin-tensor fields
characterized by a Young tableaux with 2 rows ($n_1 \geq n_2$)
\begin{equation}\label{Young k2}
\Phi_{(\mu)_{n_1},(\nu)_{n_2}}(x) \equiv
\Phi_{\mu_1\ldots\mu_{n_1},\nu_1\ldots\nu_{n_2}}(x)
\longleftrightarrow
\begin{array}{|c|c|c c c|c|c|c|c|c| c| c|}\hline
  \!\mu_1 \!&\! \mu_2\! & \cdot \ & \cdot \ & \cdot \ & \cdot\ & \cdot\ & \cdot\  & \cdot\ &
  \cdot\
  & \cdot\    &\!\! \mu_{n_1}\!\! \\
   \hline
    \! \nu_1\! &\! \nu_2\! & \cdot\
   & \cdot\ & \cdot & \cdot & \cdot & \cdot & \cdot & \!\!\nu_{n_2}\!\!   \\
  \cline{1-10}
\end{array}\ .
\end{equation}
The field $\Phi_{(\mu)_{n_1},(\nu)_{n_2}}(x) $ is symmetric with
respect to the permutations of each type of its indices $(\mu)_{n_1},
(\nu)_{n_2}$ and must obey the equations
\begin{eqnarray}
\label{Eq-0}
\left(\imath\gamma^{\mu}\partial_{\mu}\Phi_{(\mu)_{n_1},(\nu)_{n_2}},\
\gamma^{\mu_1} \Phi_{\mu_1\mu_2\ldots\mu_{n_1},\ (\nu)_{n_2}},\
\gamma^{\nu_1} \Phi_{(\mu)_{n_1},\nu_1\nu_2 ... \nu_{n_2}},\
\Phi_{\{(\mu)_{n_1},\nu_1\}\nu_2...\nu_{n_2}}\right)=0.
\end{eqnarray}
After the introduction of 2 pairs of creation and annihilation
operators, we have
\begin{eqnarray}\label{comrels}
[a^i_\mu, a_\nu^{j+}]=-\eta_{\mu\nu}\delta^{ij}\,, \qquad
\delta^{ij} = diag(1,1)\,, \eta_{\mu\nu} = diag(1,-1,\ldots,-1)\,.
\end{eqnarray}
The general (Dirac-like spinor) state in this Fock space
$\mathcal{H}^2 = \mathcal{H}_1 \bigotimes \mathcal{H}_2$ has the
form
\begin{eqnarray}
\label{PhysState1} |\Phi\rangle &=&
\textstyle\sum_{n_1=0}^{\infty}\sum_{n_2=0}^{n_1}\Phi_{(\mu)_{n_1},(\nu)_{n_2}}(x)\,
a^{+\mu_1}_1\ldots\,a^{+\mu_{n_1}}_1a^{+\nu_1}_2\ldots\,a^{+\nu_{n_2}}_2|0\rangle
.
\end{eqnarray}
The deduction of Eqs. (\ref{Eq-0}) proceeds by means of
the operators $t_0, t^i, t_{12}$ as follows:
\begin{eqnarray}
{t}_0|\Phi\rangle= -
i\tilde{\gamma}^{\mu}\partial_\mu|\Phi\rangle=0\,, \quad
{t}^i|\Phi\rangle= \tilde{\gamma}^{\mu}a^i_\mu|\Phi\rangle = 0\,,
  \quad t_{12}|\Phi\rangle = a^{+}_{1{}\mu} a^{\mu}_2|\Phi\rangle =0\footnotemark \label{t}
\end{eqnarray}\footnotetext{As opposed to the AdS space,
the corresponding HS symmetry superalgebra of all the operators,
including those of (\ref{t}), is a Lie superalgebra.}

Now, we can generalize this construction to spin-tensors
corresponding to a $k$-row ($k \leq [(d-1)/2]$) Young tableaux. To
this end, one should introduce a Fock space   $\mathcal{H}^k =
\mathcal{H}_1 \bigotimes \ldots \bigotimes \mathcal{H}_k$ with $k$
pairs of $a_\mu^{j+}, a^i_\mu$ and introduce operators
(\ref{t}), this time, however, with $i,j = 1, \ldots, k$ for $t_{ij} =
a^{+}_{i{}\mu} a^{\mu}_j, i < j$. Then, an LF can be found
partially according to the above-developed  principles
\cite{flatfermmix}. The program of an LF construction on the basis
of this method for both massive and and massless fermionic HS
fields with a two-row Young tableau was realized in
\cite{flatfermmix}.

 \vspace{-1ex} \noindent
\paragraph{4. Summary} In this article, we have briefly considered
the construction of an LF for free massive and massless HS fields on a
basis of the BFV--BRST approach. In addition, note, that the value
of reducibility stage for a gauge  LF increases  with
the number of rows of the Young tableaux. Second, there exists
a possibility to eliminate the set of algebraic second-class constraints
from the system of all constraints so as to reduce the amount of
calculations and the form of the final LF, however, with an appearance
of some off-shell algebraic conditions (such as tracelessness or
$\gamma$-tracelessness). Third, it is necessary to realize a SUSY
generalization of LF to HS fields, which will permit one to construct,
on the basis of the research \cite{BuchbinderTsulaia} of interacting
bosonic  HS fields, an interacting theory with fermionic HS fields.

\vspace{-1ex} \noindent
\paragraph{Acknowledgements} The author
thanks the organizers of the SQS'07 Workshop (JINR, Dubna, Russia)
for support and hospitality.

\end{document}